\documentclass[aps,prl,twocolumn,english,superscriptaddress]{revtex4-1}
\usepackage{amsmath,amssymb,amsfonts}
\usepackage{CJK}
\usepackage{bm}
\usepackage{tikz}
\usepackage{babel}
\usepackage{color}
\usepackage{graphicx}

\usepackage{xspace}
\usepackage{epstopdf}
\usepackage{dcolumn}
\usepackage{longtable}
\usepackage{multirow}

\def \L {\mathcal{L}}

\def \k {\mathbf{k}}

\def \i {\hat{i}}

\def \Z {\mathbb{Z}}

\def \R {\mathbb{R}}

\def \G {\bm{G}}
\def \e {\bm{e}}

\def \btau {\bm{\tau}}
\def \bomega {\bm{\omega}}

\def \Hom {\mathrm{Hom}}

\def \Ext {\mathrm{Ext}}

\def \K {\bm{K}}

\def \t {\mathbf{t}}
\def \i {\mathrm{i}}
\def \L {\mathsf{L}}

\def \i {{\mathrm{i}}}

\def \Z {\mathbb{Z}}

\def \k {\bm{k}}

\def \t {\bm{t}}
\def \i {\mathrm{i}}

\begin{document}

\title{Unified Theory of Quantum Crystalline Symmetries
}

\author{Y. X. Zhao}
\email[]{zhaoyx@nju.edu.cn}
\affiliation{National Laboratory of Solid State Microstructures and Department of Physics, Nanjing University, Nanjing 210093, China}
\affiliation{Collaborative Innovation Center of Advanced Microstructures, Nanjing University, Nanjing 210093, China}
\author{L. B. Shao}
\affiliation{National Laboratory of Solid State Microstructures and Department of Physics, Nanjing University, Nanjing 210093, China}
\affiliation{Collaborative Innovation Center of Advanced Microstructures, Nanjing University, Nanjing 210093, China}

\begin{abstract}
	Symmetry groups are \textit{projectively} represented in quantum mechanics, and crystalline symmetries are fundamental in condensed matter physics.
	Here, we systematically present a unified theory of quantum mechanical space groups from two complementary aspects. First, we provide a decomposition form for the space-group factor systems to characterize all quantum space groups. It consists of three factors, the factor system for the translation subgroup $L$, an in-homogeneous factor system for the point group $P$, and a factor connecting $L$ and $P$.
	The three factors satisfy three consistency equations, which are exactly solvable and can completely exhaust all factor systems for space groups. 
	Second, since factors systems are classified by the second cohomology group, we show the (co)homology groups for space groups can be derived from Borel's equivariant (co)homology theory, which leads to an algorithm that can compute all (co)homology groups for space groups.
	To demonstrate the general theory, we explicitly present quantum wallpaper groups with the $\Z_2$ gauge group. Furthermore, as a primitive application, we find the time-reversal invariant quantum space groups with inversion symmetry can lead to a novel clifford band theory, where each band is fourfold degenerate to represent certain real Clifford algebras with topologically nontrivial pinor structures over the Brillouin zone. Our work serves as a foundation for exploring quantum mechanical space groups, and can find applications in spin liquids, unconventional superconductors, and artificial lattice systems, including cold atoms, photonic and phononic crystals, and even LC electric  circuit networks. 
\end{abstract}
\maketitle

\textcolor{blue}{\textit{Introduction}}
Symmetry principle is a main pillar of physics, especially for quantum mechanics~\cite{Weyl-book}. An essential distinction of quantum mechanical symmetry from classical symmetry is that the symmetry group is \textit{a priori} projectively represented, and the group structure is extended by the intrinsic $U(1)$ gauge group of quantum states.
In high energy physics, the most fundamental group is the Poincar\'{e} group, where projective representations correspond to Weyl, Majorana and Dirac spinors, and the ordinary ones give rise to gauge bosons, such as electromagnetic fields~\cite{Wigner-Lorentz,Bargmann-Wigner}.
For non-relativistic physics, time and space are separated, and space groups serve as defining symmetries for condensed matter systems.  In this article, we focus on the quantum version of the space groups, namely the extended space group by the $U(1)$ gauge group. 
Compared with high energy physics, there are a variety of space groups (230 in three dimensions), while the Poincar\'{e} group is unique for high energy physics. Moreover, a given space group usually corresponds to more complex quantum versions. In this respect, condensed matter is able to exhibit much richer quantum forms of matter, compared with high energy physics. 




Analogous to classical space groups that have been instrumental in condensed matter physics~\cite{dresselhaus,Space-Time-Crystal,Zhang2019,Vergniory2019,Tang2019a}, quantum space groups ubiquitously govern physical properties of various systems. In principle, each energy eigenspace of any correlated quantum many-body system should represent a quantum space group. A more practical scenario is to consider quantum particle states on a lattice coupled with certain gauge flux configurations, namely tight-binding models with gauge fields~\cite{Anderson-emergent_Gauge,Lieb-Theorem,Wen-PSG,Kitaev2006,Zhao_Trans}, which describe a wide range of physical systems even beyond condensed matter.  For instance, in the mean-field theories or exactly solvable models for spin liquids, there are emergent gauge fields, and projective space groups have been proposed as a classification scheme for quantum orders~\cite{Anderson-emergent_Gauge,Wen-PSG,Kitaev2006,Xiao-Gang_RMP,Zhao_Second_SL}. For crystalline superconductors, the Bogoliubov quasiparticles are coupled with the remaining $\Z_2$ gauge fields after the $U(1)$ symmetry breaking due to the condensation of cooper pairs with charge $2e$~\cite{Sigrist-RMP,Krauss_Wilczek_Discrete_Gauge,Discrete_Gauge_Theory}. Meanwhile, artificial systems have been used to simulate tight-binding models with gauge fields. For instance, cold atoms in optical lattices can realize hopping amplitudes with arbitrary phases~\cite{Optical_Lattice_RMP,Optical_Lattice_2,ColdAtom_Phys-Today,Zhu_ColdAtom_Review}. For photonic and phononic crystals, with natural time-reversal ($T$) symmetry, the $\Z_2$ gauge fields can be simulated~\cite{Ozawa2019rmp,MaGuancong2019nrp}. Recently, LC electric networks have also been used to simulate tight-binding models with $\Z_2$ gauge fields for topological phases~\cite{Ronny-Topolectric,Ronny2019JulPRL,Yu_4D_QHE}. 

Although the obvious fundamental importance, a theoretical foundation is still absent in the literature, and it is the aim of this Letter to fill the gap. First, we show that each factor system for a quantum space group is gauge equivalent to a decomposed form with three components. 
The three components satisfy three consistency equations, which are practically solvable and therefore can exhaust all factor systems. Second, as the classification of factor systems is given by $H^2(G,A)$, we show that all (co)homology groups $H^{n}(G,A)$ for any space group $G$ with gauge group $A$, can be derived from equivariant (co)homology theory based on the Borel construction~\cite{ATIYAH-Equivariant_Cohomology,Brown-book}, which produces an efficient algorithm to compute (co)homology groups for all space groups.
The above two aspects are complementary: For physical applications, it is indispensable to present concrete factor systems, while the classification group $H^{2}(G,A)$ can help to organize numerous gauge-equivalent factor systems.
Based on the established foundation, we explicitly work out the quantum wallpaper groups with gauge group $\Z_2$. As a primitive application, we find a novel Clifford band theory for quantum space groups with inversion symmetry ($P$), where each band is multiply degenerate to represent real Clifford algebras with topological nontrivial pinor structures~\cite{Clifford_module}.

\textcolor{blue}{\textit{The decomposition of factor systems}}
We start with recalling basics of a projective representation $V$ of a group $G$. For  $g_1,g_2\in G$, the combined operator $V_{g_1}V_{g_2}$ and the operator $V_{g_1g_2}$ should transform an aribitrary $|\psi\rangle$ to the the same state. But, because of the intrinsic $U(1)$ gauge symmetry of quantum states, in general $V_{g_1g_2}$ is equal to $V_{g_1}V_{g_2}$ only up to a phase factor $\nu(g_1,g_2)$, namely
\begin{equation}
V_{g_1}V_{g_2}=\nu(g_1,g_2)V_{g_1g_2}.
\end{equation}
From the associativity of linear operators, the factor system $\nu$ satisfies 
\begin{equation}\label{2-cocycle_Eq}
\nu(g_1,g_2)\nu(g_1g_2,g_3)=\nu(g_1,g_2g_3)\nu(g_2,g_3).
\end{equation}
As $T$ inverses a phase factor, if $\nu$ is valued in $\Z_2\in\{\pm 1\}$, $T$ is preserved. Below, we consider $\nu$ is valued in a subgroup $A$ of $U(1)$, and are especially interested in $A=\Z_2$ or $U(1)$.
Redefining the phases of each $V_{g}$ by a function $\chi$ from $G$ to $A$, we obtain another equivalent projective representation $\tilde{V}_g=\chi(g)V_g$, and therefore the equivalence relation for factor systems is given by
\begin{equation}\label{Equivalence}
\nu(g_1,g_2)\sim \nu(g_1,g_2)\frac{\chi(g_1g_2)}{\chi(g_1)\chi(g_2)}.
\end{equation}

If $G$ is a space group, there are infinite number of group elements, which makes it difficult to solve Eq.~\eqref{2-cocycle_Eq} and identify equivalent solutions by \eqref{Equivalence}. By generalizing theorem 9.4 in Ref.~\cite{mackey1958}, we find that any equivalence class of factor systems contain a representative with the decomposed form~\cite{a-Note_Decomposition},
\begin{equation}\label{Decomposition}
\begin{split}
\nu(g_1,g_2)=&\sigma(\t_1,R_1\t_2)\sigma(\t_1+R_1\t_2,\bomega(R_1,R_2))\\& g^{-1}(R_1\t_2,R_1)\alpha(R_1,R_2).
\end{split}
\end{equation}
The proof can be found in the supplemental material (SM)~\cite{Supp}. Here, an space group element is denoted as $g=\{\t|R\}$, with $\t$ a lattice translation vector and $R$ an element of the point group $P$. The group multiplication is then given as $g_1g_2=\{\t_1+R_1\t_2+\bomega(R_1,R_2)|R_1R_2\}$, where $\bomega(R_1,R_2)$ is a lattice translation vector~\cite{b-Note_Multiplication}.
Here, $\sigma$ is the factor system for the translational subgroup $L\cong\Z^d$ of $G$, obtained from restricting $G$ onto $L$, namely
$\sigma(\t_1,\t_2)=\nu(\{\t_1|E\},\{\t_2|E\})$. And $\alpha(R_1,R_2)=\nu(\{0|R_1\},\{0|R_2\})$. It noteworthy that $\alpha$ is not a factor system for $P$, unless $G$ is symmorphic.

The components of Eq.~\eqref{Decomposition} are required to satisfy the following consistency equations:
\begin{equation}\label{Flux-condition}
\frac{\sigma(R^{-1}\t_1,R^{-1}\t_2)}{\sigma(\t_1,\t_2)}=\frac{g(\t_1+\t_2,R)}{g(\t_1,R)g(\t_2,R)},
\end{equation}
\begin{equation}\label{Covariance_g}
\frac{g(\t,R_1R_2)}{g(\t,R_1)g(R_1^{-1}\t,R_2)}=\frac{\sigma(\bomega(R_1,R_2),\t)}{\sigma(\t,\bomega(R_1,R_2))},
\end{equation}   
\begin{equation}\label{Point_group_factor}
\begin{split}
\frac{\alpha(R_1,R_2)\alpha(R_1R_2,R_3)}{\alpha(R_1,R_2R_3)\alpha(R_2,R_3)}=g^{-1}(R_1\bomega(R_2,R_3),R_1)\\
\times \frac{\sigma(R_1\bomega(R_2,R_3),\bomega(R_1,R_2R_3))}{\sigma(\bomega(R_1,R_2),\bomega(R_1R_2,R_3))}.
\end{split}
\end{equation}
which are derived in detail in the SM~\cite{Supp}. The three equations are sufficient for the decomposition \eqref{Decomposition} to satisfy Eq.~\eqref{2-cocycle_Eq}. To preserve the decomposed form \eqref{Decomposition}, $\chi(g)$ in the equivalence relation \eqref{Equivalence} must take the decomposed form $\chi(g)=\psi(\t)\varphi(R)$. Then, \eqref{Equivalence} leads to the following equvalence relations for components~\cite{Supp}:
\begin{eqnarray}
\sigma(\t_1,\t_2) &\sim& \sigma(\t_1,\t_2)  \frac{\psi(\t_1+\t_2)}{\psi(\t_1)\psi(\t_2)},\label{sigma-equivalence}\\
g(\t,R) &\sim& g(\t,R)\frac{\psi(R^{-1}\t)}{\psi(\t)},\label{g-equivalence}\\
\alpha(R_1,R_2) &\sim& \alpha(R_1,R_2)\frac{\psi(\bomega(R_1,R_2))\varphi(R_1R_2)}{\varphi(R_1)\varphi(R_2)}\label{alpha-equivalence}.
\end{eqnarray}
We observe that \eqref{sigma-equivalence} is just the usual equivalence relation for $L$. The equivalence relation \eqref{alpha-equivalence} for $\alpha$ is not the usual equivalence relation for factor systems of $P$, but is modified by $\psi$ with the $\bomega$-twist.

It is noteworthy several general features of the consistency Eqs.~\eqref{Flux-condition}, \eqref{Covariance_g} and \eqref{Point_group_factor}. For a fixed $R$ in Eq.~\eqref{Flux-condition}, the right-hand side is just a trivial factor system for $L$. Hence, $\sigma$ must be compatible with the point group $P$, i.e., it is still in the same equivalence class after a transformation by $R$. Conversely, with given $\sigma$ and a particular solution $g_0$ for $g$ , the general solution to  Eq.~\eqref{Flux-condition} is $g=g_0(\t,R) e^{\i\k(R)\cdot \t}$, since $g/g_0$ is multiplicative for $\t$. For $A=\Z_2$, $\k$ is valued in inversion invariant points in the first Brillouin zone (BZ).

The consistency Eqs.~\eqref{Covariance_g} and \eqref{Point_group_factor} can be regarded as inhomogeneous equations with sources from the $\bomega$-twist. For symmorphic groups with $\bomega=0$, they are homogeneous, and $\alpha$ is just a factor system for $P$, since $P$ is a subgroup of symmorphic $G$. For nonsymmorphic space groups, the product of a particular solution $(g_0,\alpha_0)$ with any homogeneous solution is still a solution to Eqs.~\eqref{Covariance_g} and \eqref{Point_group_factor}.

As the factor systems $\sigma$ for translation groups are well known, the consistency equations can be reduced to a finite set of linear equations, and therefore can be exactly solved to obtain all factor systems for all space groups.



\begin{table*}
	\begin{tabular}{c|ccccccccccccccccc}
		$G$ & $p1$ & $p2$ & $pm$ & $pg$ & $cm$ & $pmm$ & $pmg$ & $pgg$ & $cmm$ & $p4$ & $p4m$ & $p4g$ & $p3$ & $p3m1$ & $p31m$ & $p6$ & $p6m$ \\
		\hline
		$H^2(G,\Z_2)$ & $\Z_2$ & $\Z_2^4$ & $\Z_2^4$ & $\Z_2$ & $\Z_2^2$ & $\Z_2^8$ & $\Z_2^4$ & $\Z_2^2$ & $\Z_2^5$ & $\Z_2^3$ & $\Z_2^6$ & $\Z_2^3$ & $\Z_2$ & $\Z_2^2$ & $\Z_2^2$ & $\Z_2^2$ & $\Z_2^2$\\
		$H^2(G,U(1))$ & $U(1)$ & $U(1)$ & $\Z_2^2$ & $0$ & $\Z_2$ & $\Z_2^4$ & $\Z_2$ & $0$ & $\Z_2^2$ & $U(1)$ & $\Z_2^3$ & $\Z_2$ & $U(1)$ & $\Z_2$ & $\Z_2$ & $U(1)$ & $\Z_2^2$ 
	\end{tabular}
	\caption{Classification tables of quantum wallpaper groups with gauge group $\Z_2$ and $U(1)$.\label{tab:cls_tab}}
\end{table*}

\textcolor{blue}{\textit{Cohomology groups of space groups}}
Gauge equivalence classes of quantum versons for a given space group $G$ form the second group cohomology $H^2(G,A)$, which is just the gauge equivalence classes of factor systems under multiplication. It is not easy to solve $H^2(G,A)$ directly from the algebraic definitions [Eqs.~\eqref{2-cocycle_Eq} and \eqref{Equivalence}]. We transform it into a topological problem. The group cohomology $H^n(G,A)$ is equivalent to the topological cohomology $\mathcal{H}^n(BG,A)$~\cite{Dijkgraaf-Witten}, namely
$
H^n(G,A)\cong \mathcal{H}^n(BG,A),
$
where $BG$ is the so-called classifying space of the group $G$. If we have a contractible space $EG$ with a free $G$-action, then the orbital space $EG/G$ is a classifying space $BG$ for $G$.


Probably the most elementary example is given by the translation group $L\cong\Z^d$ of a $d$D lattice, which acts freely on the linear space $\R^d$. Therefore, $E\Z^d=\R^d$, and $B\Z^d=\R^d/\Z^d\approx T^d$. Hence, the classification of factor systems for the translational group is given by
\begin{equation}\label{L-cohomo}
H^2(\Z^d,A)\cong \mathcal{H}^2(T^d,A)\cong A^{d(d-1)/2}.
\end{equation}

For an arbitrary space group $G$,
the contractible total space $EG$ can be chosen as 
\begin{equation}
EG=\R^d\times EP.
\end{equation}
Here, $EP$ is a contractible space with a free $P$-action, which can be systematically constructed~\cite{Milnor_Construction}.
Now, $\R^d$ is naturally a $G$-space with the $G$-action, but the $G$ action is in general not free. Meanwhile, the free action of $P$ on $EP$ simply induces a $G$-action, with the translations acting trivially on $EP$. Hence, the diagonal $G$-action on $\R^d\times EG$ is given by
$\{\t|R\}(\bm{r},p)=(R\bm{r}+\t+\btau(R), Rp)$.
where $\bm{r}\in \R^d$ and $p\in EP$, and $\btau(R)$ is the fractional translation associated with $R$~\cite{b-Note_Multiplication}. It is obvious that the diagonal action is a free $G$-action over $EG=\R^d\times EP$.
Hence, we can choose the classifying space as 
\begin{equation}
BG=\R^d\times_G EP=T^d\times_P EP.
\end{equation}
Here, $\R^d\times_G EG=(\R^d\times G)/G$. In the second equality, the space $\R^d\times_G EG$ is simplified by factoring out the action of the translation subgroup $\Z^d\subset G$, which reduces $\R^d$ to $T^d$. Accordingly, the $P$-action on $T^d$ is specified by $\btau$. Then, the diagonal action of $P$ on $T^d\times EP$ is given by $R(\bm{r},p)=(R\bm{r}+\btau(R),Rp)$, with the first component modulo integer translation vectors.
Hence, the cohomology groups are given by
\begin{equation}
H^n(G, A)\cong \mathcal{H}^n(T^d\times_P EP,A).
\end{equation}
Here, $\mathcal{H}^n(T^d\times_P EP,A)$ is referred to as the equivariant cohomology group  $\mathcal{H}^n_P(T^d,A)$, and the construction above is called the Borel construction. Then, following the standard procedure from equivariant (co)homology~\cite{Brown-book}, a total (co)chain complex can be constructed from the (co)chain complexes for $T^d$ and $EP$, and therefore the (co)homology groups can be derived. The algorithm is demonstrated by wallpaper group $pg$ in detail in the SM~\cite{Supp}, and the classification tables for quantum wallpaper groups are presented in Tab.\ref{tab:cls_tab} for both gauge groups $\Z_2$ and $U(1)$.

\textcolor{blue}{\textit{Translation groups}} Since all space groups have translation subgroups $L\cong\Z^d$, we first introduce factor systems $\sigma$ for $L$. 
Let us consider the factor systems in the form,
\begin{equation}\label{L-factor}
\sigma(\t_1,\t_2)=\exp(-\pi \i~\sum_{ij}t^{i}_1 A_{ij} t^j_2).
\end{equation}
Here, $t$ is is a vector of integers given by $\t=t^i\e_i$ with $\e_i$ unit translation vectors. 
For $A=U(1)$, the matrix elements $A_{ij}$ are valued in $[-1,1)$, and
for $\Z_2=\{\pm 1\}$, $A_{ij} \in \{0,1\}$. To present all factor system classes given by \eqref{L-cohomo} , we may reduce the matrix $A$ to a canonical form, namely a lower-triangular matrix with vanishing diagonal entries. Then, each of the $d(d-1)/2$ lower-triangular entries contributes a generator for $A^{d(d-1)/2}$ in \eqref{L-cohomo}.

The factor system \eqref{L-factor} physically corresponds to the gauge fluxes through unit faces on a lattice. Let $\L$ be a projective representation with $\sigma$, then
$\L_{\t_1}\L_{\t_2}=\sigma(\t_1,\t_2)\L_{\t_1+\t_2}$, which leads to the Wilson loop operator $W(\t_1,\t_2)=\L_{\t_1}\L_{\t_2}\L_{\t_1}^{-1}\L_{\t_2}^{-1}=\sigma(\t_1,\t_2)/\sigma(\t_2,\t_1)$. 
In the canonical form, $A_{ij}$ with $i>j$ is just the flux through the parallelogram spanned by $\e_i$ and $\e_j$.

\textcolor{blue}{\textit{Wallpaper groups with gauge group $\Z_2$}} We now address several key points for quantum wallpaper groups with the $T$-invariant $\Z_2$ gauge group~\cite{c-Note_Multiplication}. As aforementioned, for symmorphic space groups, $\alpha$ is just the factor system for the point group corresponding to $H^2(P,A)$. Therefore, we only need to solve Eqs.~\eqref{Flux-condition} and \eqref{Covariance_g} with equivalence relations \eqref{sigma-equivalence} and \eqref{g-equivalence}. This is particularly easy for wallpaper groups with $A=\Z_2$. First, $\sigma$ has only two possibilities with flux $\pi$ or $0$ through the fundamental domain. Second, $g=\Delta(\t,R)e^{\i\k(R)\cdot\t}$ has only $4^{N_P}$ possibilities with $N_P$ the number of generators for $P$, since $\k(R)$ values in the four inversion invariant points $\K_a\in \Z_2\times\Z_2$ in the BZ.

For nonsymmorphic groups, all three consistency equations should be simultaneously considered. There are four nonsymmorphic wallpaper groups: $pg$, $pmg$, $pgg$ and $p4g$. We solve factor systems for them all with $A=\Z_2$, and find two remarkable features. First, for all of them, $\sigma$ is trivial for translation subgroup. This is because the nontrivial $\sigma$ contradicts with a glide reflection in Eq.~\eqref{Covariance_g} (see the SM for details \cite{Supp}), but all nonsymmorphic wallpaper groups contain glide reflections. Second, multiplicative $\psi$ with $\psi(\t_1+\t_2)=\psi(\t_1)\psi(\t_2)$, although does not change $\sigma$, can impose equivalence relations among $(g,\alpha)$'s.
With $\sigma=1$, we then solve the homogeneous \eqref{Covariance_g} and the inhomogeneous \eqref{Point_group_factor} with the source term solely from $g$, and the results are tabulated in the SM~\cite{Supp}. Then, we should use the gauge equivalence relations \eqref{g-equivalence} and \eqref{alpha-equivalence} with $\psi_a(\t)=e^{\i \K_a\cdot\t}$ to further reduce the number of equivalence classes to obtain representatives for $H^{2}(G,\Z_2)$.


\textcolor{blue}{\textit{The Clifford band theory}} We now present a primitive application of quantum space groups, namely the Clifford band theory for quantum space groups with $PT$ symmetry. Without loss of generality, let us consider the $2$D rectangular lattice. $T$-invariance requires the flux through each plaquette be $0$ or $\pi$, and we specialize in the latter, which leads to
\begin{equation}
\{\L_x,\L_y\}=0,
\end{equation}
for two unit translation operators $\L_x$ and $\L_y$.
Observing that
$[\L_x^2,\L_y]=0$ and $[\L_y^2,\L_x]=0$,
we can represent the generators in momentum space as $\mathcal{L}_x$ and $\mathcal{L}_y$ with 
\begin{equation}
\{\mathcal{L}_x(\k),\mathcal{L}_y(\k)\}=0,~~ \mathcal{L}_{x,y}(\k)^2=e^{\i k_{x,y}}.
\end{equation}
Locally in the BZ, there is a unique irreducible representation for $\mathcal{L}_x$ and $\mathcal{L}_y$~\cite{Zhao_Trans}, which is given by
\begin{equation}
\mathcal{L}_i=e^{\i k_i/2}\sigma_i 
\end{equation} 
with $i=x,y$. 
To see this, let us renormalize the translators as $\hat{\mathcal{L}}_{x,y}=e^{-\i k_{x,y}/2}\mathcal{L}_{x,y}$, which satisfy the Clifford algebra,
\begin{equation}
\{\hat{\mathcal{L}}_x(\k),\hat{\mathcal{L}}_y(\k)\}=0,\quad \hat{\mathcal{L}}_{x,y}(\k)^2=1.
\end{equation}
But the Clifford algebra has only one unique irreducible representation. Thus, each band is at least twofold degenerate to form a pinor structure over the Brillouin zone with the Clifford algebra represented. 

Furthermore, the momentum-dependent unitary translation operators $\mathcal{L}_{x,y}$ have nontrivial winding numbers along large circles in the Brillouin zone. The winding numbers can be derived from
\begin{equation}\label{Winding-number}
N=\frac{1}{2\pi \i}\oint dk_i~\mathrm{tr}\mathcal{L}_i^\dagger\partial_{k_i}\mathcal{L}_i,
\end{equation}
and equal $1$ for both $i=x,y$. The nontrivial winding numbers imply that the pinor structure for each band is topologically nontrivial.

Let us proceed to consider inversion symmetry $P$. The most general case is the symmorphic wallpaper group $p2$ with the point group $C_2$, since additional symmetries would impose more constraints through consistency equations. 
When restricted on the point subgroup $C_2$ , a factor system $\nu$ of $p2$ gives a factor system $\alpha$ of $C_2$. According to $H^2(C_2,\Z_2)\cong\Z_2$, we have $\alpha(P,P)=s_p$,
where $s_p=\pm 1$ corresponding to trivial and nontrivial $C_2$ factor systems, respectively. Then, the projective representation $\mathsf{P}$ of $P$ follows the relation:
\begin{equation}
\mathsf{P}^2=s_p.
\end{equation}
For $H^2(p2,\Z_2)\cong \Z_2^4$, two $\Z_2$ components are contributed by the factor systems for translation subgroup and the point group $C_2$, respectively, and the other two components come from 
\begin{equation}
g(\t,P)=(-1)^{\sum_i q_it^i},
\end{equation}
where $q_i=0$ or $1$ with $i=x,y$. 

The four solutions for $g$ dictate the commutation relations of $\mathcal{P}$ and $\hat{\mathcal{L}}_i$.
From the decomposition form \eqref{Decomposition}, it follows that
$\mathsf{P}\L_i=(-1)^{q_i}\L_i^{-1}\mathsf{P}$.
Since $\L_i^{-1}\mathsf{P}=-\L_i^{-2} \L_i\mathsf{P}$, we can derive that in momentum space,
$\mathcal{P}\mathcal{L}_i=(-1)^{q_i}e^{-\i k_i}\mathcal{L}_i\mathcal{P}$, where $\mathcal{P}=U_P\hat{I}$ with $U_P$ a unitary operator and $\hat{I}$ the momentum inversion.
In terms of the renormalized translators, $\hat{\mathcal{L}}_{i}=e^{-\i k_{i}/2}\mathcal{L}_{i}$, the above equation can be cast into
\begin{equation}
\hat{\mathcal{L}}_i\mathcal{P}=(-1)^{q_i}\mathcal{P}\hat{\mathcal{L}}_i.
\end{equation}
namely, that $\hat{\mathcal{L}}_i$ anti-commutes (commutes) with $\mathcal{P}$ if $q_i=1$ ($q_i=0$).

We further take $T$ symmetry into consideration. As an anti-unitary symmetry,  it is represented by $\mathcal{T}=U_T\hat{\mathcal{K}}\hat{I}$ in momentum space, where $U_T$ a unitary operator and $\hat{\mathcal{K}}$ complex conjugation. With the $\Z_2$ gauge group, $T$ symmetry is naturally preserved, and therefore spatial operators commute with $\mathcal{T}$. Then, we have the algebraic relations,
\begin{equation}
[\mathcal{T},\mathcal{P}]=0,~~ [\mathcal{T},\hat{\mathcal{L}}_{x,y}]=0,~~\{\i,\mathcal{T}\}=0,~~\mathcal{T}^2=s_t.
\end{equation}
Here,  $s_t=1$ for particles with integer spins or half-integer spins, like electrons, without spin-orbital coupling (SOC). $s_t=-1$ for particles with half-integer spins and SOC~\cite{Schnyder2008}. Since $\mathcal{T}$ is anti-unitary, we have included the imaginary unit $\i$ as an operator. Note that the renormalization $\hat{\mathcal{L}_i}=e^{-\i k_i/2}\mathcal{L}_i$ does not affect the commutativity of $\mathcal{T}$ with translation operators.

\begin{table}
	\begin{tabular}{c|cccc|cccc}
		\hline
		{} & \multicolumn{4}{c}{$s_ts_p=+$} & \multicolumn{4}{c}{$s_ts_p=-$}\\
		\hline
		$q_x$ & $0$ & $1$ & $0$ & $1$ & $0$ & $1$ & $0$ & $1$\\
		$q_y$ & $0$ & $0$ & $1$ & $1$ & $0$ & $0$ & $1$ & $1$\\ 
		\hline
		$C^{n,m}$ & $(2,2)$ & $(1,3)$ & $(1,3)$ & $(0,4)$ & $(4,0)$ & $(3,1)$ & $(3,1)$ & $(2,2)$\\
		D  & 2 & 2 & 2 & 4 & 4 & 4 & 4 & 2\\
		\hline
	\end{tabular}
	\caption{The real Clifford algebras of operators. The last row lists the unique dimension of irreducible representations of the corresponding real Clifford alegbra. \label{tab:Clifford}}
\end{table}

The complete set of generators, $\mathcal{P}$, $\mathcal{L}_{x,y}$, $\mathcal{T}$ and $\i$, can be recombined into the real Clifford algebra generators~\cite{Furusaki_Clifford}:
\begin{equation}\label{Cl-generators}
\mathcal{P}\mathcal{T},\quad  \i\mathcal{P}\mathcal{T},\quad  \i^{1-q_x}\hat{\mathcal{L}}_x, \quad \i^{1-q_y}\hat{\mathcal{L}}_y,  
\end{equation}
which anti-commute with each other, and individually satisfy
$(\mathcal{P}\mathcal{T})^2=s_ps_t$, $(\i\mathcal{P}\mathcal{T})^2=s_ps_t$, $\hat{\mathcal{L}}_x^2=(-1)^{1-q_x}$, and $\hat{\mathcal{L}}_y^2=(-1)^{1-q_y}$.
We denote a real Clifford algebra as $C^{n,m}$, where $n$ ($m$) is the number of negative (positive) generators. Algebraically, there are only eight stably non-equivalent real Clifford algebra~\cite{Atiyah-KR}, which can be represented by $C^{n,0}$ with $n=1,2,\cdots,8$. All of them have a unique irreducible representation, except that $C^{4,0}$ has two irreducible representations with the same dimension~\cite{Clifford_module,Zhao2016a,Zhao2017}. Hence, the dimension of an irreducible representation for a real Clifford algebra is unique. All possible real Clifford algebras by the generators \eqref{Cl-generators} are tabulated in Tab.\ref{tab:Clifford}.

We observe from Tab.\ref{tab:Clifford} that for a half of the possibilities, each energy band is fourfold degenerate and form a $4$D Dirac pinor structures representing the corresponding Clifford algebra. For instance, in the case of $s_ts_p=+1$ and $q_x=q_y=1$, the operators can be represented by
\begin{equation}
\mathcal{L}_x=e^{\i k_x/2}\sigma_1\otimes\tau_0,~~ \mathcal{L}_y=e^{\i k_y/2}\sigma_3\otimes \tau_0,
\end{equation}
$\mathcal{P}=\sigma_2\otimes\tau_2 \hat{I}$, and $\mathcal{T}=\hat{\mathcal{K}}\hat{I}$.
The fourfold band degeneracy goes beyond conventional band theories, for which the maximum degeneracy is twofold and originated from $PT$-invariant SOC. Moreover, the Dirac pinor structure is topologically nontrivial because of \eqref{Winding-number}.


{\color{blue}\textit{Summary and discussions}} In summary, we have presented a unified theory for quantum space groups, including a solvable decomposition form for factor systems and an algorithm for cohomology groups. As a primitive application, we show $\Z_2$ projective representations of space group with inversion symmetry can lead to the Clifford band theory. Based on the theoretical foundation established in this Letter, there are much more waiting to be explored in the quantum regime of space groups. A promising direction is to explore quantum-crystalline topological phases, as classical crystal symmetries have recently been shown to give rise to thousands of topological materials~\cite{Tang2019a,Vergniory2019,Zhang2019}. 


\bibliographystyle{apsrev4-1}
\bibliography{QSG_Ref}

\begin{thebibliography}{43}
\bibitem [{b-N()}]{b-Note_Multiplication}%
  \BibitemOpen
  \href@noop {} {}\bibinfo {note} {An equivalent way to describe a space group
  is to specify a fractional translation $\btau(R)$ for each $R\in P$, because
  $H^2(G,\Z^d)\cong H^2(G,U(1)^d)$~\cite{Crystal_Geometry}. Then,
  $\bomega(R_1,R_2)=\btau(R_1)+R_1\btau(R_2)-\btau(R_1R_2)$.}\BibitemShut
  {Stop}%
\bibitem [{c-N()}]{c-Note_Multiplication}%
  \BibitemOpen
  \href@noop {} {}\bibinfo {note} {The quantum wallpaper groups with the $\Z_2$
  gauge group all have physical interpretations, which deserve a separate
  forthcoming paper to elucidate.}\BibitemShut {Stop}%
\end{thebibliography}

\clearpage
\newpage

\onecolumngrid
\appendix

\section{Supplemental Material for ``Unified Theory of Quantum Space Groups"}
\section{Decomposition of factor systems}
We now present a proof for the decomposition form (4) in the main text for the factor system. For notational simplicity, we abbreviate $\{\t|R\}$ as $tR$, and sometimes $\{\t|E\}$ simply as $t$, when there is no confusion from the context.
Since $\{\t|R\}=\{\t|E\}\{0|R\}$, for an arbitrary projective representation $V'$ with a factor system $\nu'$, we have the identity,
\begin{equation}
V'_{tE}V'_{0R}=\nu'(tE,0R)V'_{tR}.
\end{equation}
Then, we can define an equivalent projective representation $V$ as
\begin{equation}\label{Decomposition_Gauge}
V_{tR}=\nu'(tE,0R)V'_{tR},
\end{equation}
which has the nice property:
\begin{equation}\label{Reduced_canonical_product}
V_{tE}V_{0R}=V_{tR}.
\end{equation}
We denote the factor system of $V$ as $\nu$. When $\nu$ is restricted in the translational subgroup $\Z^d$, the corresponding factor system for $\Z^d$ is denoted as $\sigma$. Then, there is the identity,
\begin{equation}\label{nu-sigma}
\nu(tE,\{\bomega(R_1,R_2)|R_1R_2\})=\sigma(\t,\bomega(R_1,R_2)),
\end{equation}
which shall be used later. It can be proved by observing that
\begin{equation}
\begin{split}
\nu(tE,\bomega(R_1,R_2)R_1R_2)V_{\{t+\bomega(R_1,R_2)|R_1R_2\}} &=V_{tE}V_{\bomega(R_1,R_2)(R_1R_2)}\\
&=V_{t}V_{\bomega(R_1,R_2)}V_{0(R_1R_2)}\\
&=\sigma(\t,\bomega(R_1,R_2))V_{t+\bomega(R_1,R_2)}V_{0(R_1R_2)}\\
&=\sigma(\t,\bomega(R_1,R_2))V_{\{t+\bomega(R_1,R_2)|R_1R_2\}},
\end{split}
\end{equation}
where the identity \eqref{Reduced_canonical_product} was used in the second equality and the last.

To decompose the factor system $\nu$, we consider the derivations of the product:
\begin{equation*}
\begin{split}
V_{t_1E}V_{0R_1}V_{t_2E}V_{0R_2}&=V_{t_1R_1}V_{t_2R_2}\\
&=\nu(t_1R_1,t_2R_2)V_{(t_1R_1)(t_2R_2)}\\
&=\nu(t_1R_1,t_2R_2)V_{(t_1E)(0R_1)t_2E(0R_1)^{-1}(0R_1)(0R_2)}\\
&=\nu(t_1R_1,t_2R_2)V_{{t_1E}(R_1t_2E)(0R_1)(0R_2)}\\
&=\frac{\nu(t_1R_1,t_2R_2)}{\sigma(\t_1+R_1\t_2,\bomega(R_1,R_2))}V_{t_1(R_1t_2)}V_{(0R_1)(0R_2)}\\
&=\frac{\nu(t_1R_1,t_2R_2)}{\sigma(\t_1+R_1\t_2,\bomega(R_1,R_2))\sigma(\t_1,R_1\t_2)\nu(0R_1,0R_2)}V_{t_1}V_{R_1t_2}V_{0R_1}V_{0R_2},
\end{split}
\end{equation*}
where the identity \eqref{nu-sigma} was used in the fifth equality.
Hence,
\begin{equation}
\frac{\nu(t_1R_1,t_2R_2)}{\sigma(\t_1+R_1\t_2,\bomega(R_1,R_2))\sigma(\t_1,R_1\t_2)\alpha(R_1,R_2)}1_{V}=V_{0R_1}V_{t_2}V^{-1}_{0R_1}V^{-1}_{R_1t_2}.
\end{equation}
Here, we have replaced $\nu(0R_1,0R_2)$ by $\alpha(R_1,R_2)$, namely, that
\begin{equation}
\alpha(R_1,R_2)=\nu(0R_1,0R_2).
\end{equation}
The right hand side is proportional to the identity operator and depends only on $t_2$ and $R_1$.
Therefore, we introduce a function $g$ from $\Z^d\times P$ to $A$ by
\begin{equation}
g^{-1}(R_1t_2,R_1)1_{V}=V_{0R_1}V_{t_2}V^{-1}_{0R_1}V^{-1}_{R_1t_2}.
\end{equation}
Thus, the decomposition form (4) in the main text is proved.

\subsection{Equivalence relations}
For a function $\chi: G\rightarrow A$, the coboundary is given by
\begin{equation}\label{2-coboundary}
d\chi(t_1R_1,t_2R_2)=\frac{\chi(\{\t_1+R_1\t_2+\bomega(R_1,R_2)|R_1R_2\})}{\chi(t_1R_1)\chi(t_2R_2)}.
\end{equation}
To preserve the form of \eqref{Reduced_canonical_product}, we require $d\chi(tE,0R)=1$, which gives the decomposition,
$
\chi(\{\t|R\})=\chi(\{\t|E\})\chi(\{0|R\})
$.
Introducing $\psi(\t)=\chi(\{\t|E\})$ and $\varphi(R)=\chi(\{0|R\})$, $\chi$ can be expressed into
\begin{equation}\label{chi-decomposition}
\chi(\{\t|R\})=\psi(\t)\varphi(R).
\end{equation}
Substituting Eq.~\eqref{chi-decomposition} into Eq.~\eqref{2-coboundary}, we find that if $\varphi(E)=1$, $d\chi$ conforms the decomposed form (4) in the main text, which is given by
\begin{equation}
\begin{split}
d\chi(\{\t_1|R_1\},\{\t_2|R_2\})=&\frac{\psi(\t_1+R_1\t_2+\bomega(R_1,R_2))}{\psi(\t_1)\psi(\t_2)}\frac{\varphi(R_1R_2)}{\varphi(R_1)\varphi(R_2)}\\
=&\frac{\psi(\t_1+R_1\t_2)}{\psi(\t_1)\psi(R_1\t_2)}\frac{\psi(\t_1+R_1\t_2+\bomega(R_1,R_2))}{\psi(\t_1+R_1\t_2)\psi(\bomega(R_1,R_2))}\times  \frac{\psi(R_1\t_2)}{\psi(\t_2)}\times\frac{\psi(\bomega(R_1,R_2))\varphi(R_1R_2)}{\varphi(R_1)\varphi(R_2)}.
\end{split}
\end{equation}
Compared with the decomposition form (4), the equivalence relations (8), (9) and (10) in the main text are proved.

Given two factor systems $\nu$ and $\nu'$, how to judge whether they are in the same equivalence class? The answer is to follow the following algorithm. First, we redefine them according to Eq.~\eqref{Decomposition_Gauge}. Then, the two factor systems are cast into the decomposition form (4). Accordingly, the gauge transformation is decomposed as \eqref{chi-decomposition}. Second, we calculate the wilson loop $W$ for the factor systems restricted on the translational subgroup. If they have different flux configurations, the two factor systems are non-equivalent. If they are equivalent, we transform the factor systems for the translational subgroup into the canonical form (15) with $A$ a lower-triangular matrix with vanishing diagonal entries. Third, in the last step, $\psi$ is fixed to be of the form $\psi(\t)=e^{\i\pi \k\cdot\t}$, which can be used to check whether $g$ and $g'$ are equivalent by (9). 
If $g$ and $g'$ are non-equivalent, the two factor systems are nonequivalent. Otherwise, we obtain a $\psi$, and we use it to transform $(g,\alpha)$ to obtain $(g_{\psi},\alpha_{\psi})$ with $g_{\psi}=g'$. Fourth, we can check whether $\alpha$ and $\alpha'$ are equivalent according to (10). 

The case with the coefficient $A=\Z_2=\{\pm 1\}$ is particularly solvable. In the third step, $\k$ is one of $2^d$ inversion-invariant point in the Brillouin zone. In the forth step, there are only $2^{|P|-1}$ possible $\varphi$ with $|P|$ the number of element in $P$, and therefore the equivalence can be checked by brutal force.

\subsection{Consistency equations}
\subsubsection{Inhomogeneous 2-cocycle equation for $\alpha$}
Let us consider the specification $\t_1=\t_2=\t_3=0$ for the cocycle equation (3) for $\nu$ in the main text, which gives
\begin{equation}\label{alpha-step1}
\alpha(R_1,R_2)\nu(\bomega(R_1,R_2)(R_1R_2),0R_3)=\nu(0R_1,\bomega(R_2,R_3)(R_2R_3))\alpha(R_2,R_3).
\end{equation}
The second factor of the left-hand side of the equation above can be decomposed as
\begin{equation}\label{alpha_Step2}
\nu(\bomega(R_1,R_2)(R_1R_2),0R_3)=\alpha(R_1R_2,R_3)\sigma(\bomega(R_1,R_2),\bomega(R_1R_2,R_3)).
\end{equation}
This can be derived from the following derivation:
\begin{equation}
\begin{split}
&\nu(\bomega(R_1,R_2)(R_1R_2),0R_3)V_{\{\bomega(R_1,R_2)+\bomega(R_1R_2,R_3)|R_1R_2R_3\}}\\ =&V_{\bomega(R_1,R_2)(R_1R_2)}V_{0R_3}\\
=&V_{\bomega(R_1,R_2)}V_{0(R_1R_2)}V_{0R_3}\\
=&\alpha(R_1R_2,R_3)V_{\bomega(R_1,R_2)}V_{\{\bomega(R_1R_2,R_3)|R_1R_2R_3\}}\\
=&\alpha(R_1R_2,R_3)\nu(\bomega(R_1,R_2)E,\bomega(R_1R_2,R_3)R_1R_2R_3)V_{\{\bomega(R_1,R_2)+\bomega(R_1R_2,R_3)|R_1R_2R_3\}}\\
=&\alpha(R_1R_2,R_3)\sigma(\bomega(R_1,R_2),\bomega(R_1R_2,R_3))V_{\{\bomega(R_1,R_2)+\bomega(R_1R_2,R_3)|R_1R_2R_3\}}.
\end{split}
\end{equation}
For the right-hand side of Eq.~\eqref{alpha-step1}, the first factor can be decomposed according to (4) in the main text as
\begin{equation}\label{alpha_Step3}
\begin{split}
& \nu(0R_1,\bomega(R_2,R_3)(R_2R_3))\\
=& g^{-1}(R_1\bomega(R_2,R_3),R_1)\sigma(R_1\bomega(R_2,R_3),\bomega(R_1,R_2R_3))\alpha(R_1,R_2R_3).
\end{split}
\end{equation}
Substituting Eqs.~\eqref{alpha_Step2} and \eqref{alpha_Step3} into Eq.~\eqref{alpha-step1}, we can obtain (7) in the main text, which we write down again for the reader's convenience,
\begin{equation*}
\frac{\alpha(R_1,R_2)\alpha(R_1R_2,R_3)}{\alpha(R_1,R_2R_3)\alpha(R_2,R_3)}=g^{-1}(R_1\bomega(R_2,R_3),R_1)\frac{\sigma(R_1\bomega(R_2,R_3),\bomega(R_1,R_2R_3))}{\sigma(\bomega(R_1,R_2),\bomega(R_1R_2,R_3))}.
\end{equation*}

Based on (7), it is easy to check that
\begin{equation}
\alpha(E,R)=\alpha(R,E)=g(0,R)=g(t,E)=1.
\end{equation}
With $g(0,R)=1$, equation \eqref{alpha_Step2} can also be directly derived from (4) in the main text.

\subsubsection{Compatible flux condition for $\sigma$}
Let us express the cocycle equation (3) in terms of the decomposed factor system (4), which gives
\begin{equation}\label{sigma_Step1}
\frac{g(R_1\t_2,R_1)g(R_1R_2 \t_3,R_1R_2)}{g(R_1\t_2+R_1R_2\t_3+R_1\bomega(R_2,R_3),R_1)g(R_2\t_3,R_2)}=\frac{A}{B}C.
\end{equation}
Here,
\begin{multline}
A=\sigma(\t_1,R_1\t_2)\sigma(\t_1+R_1\t_2,\bomega(R_1,R_2))\\\sigma(\t_1+R_1\t_2+\bomega(R_1,R_2),R_1R_2 \t_3)\\ \sigma(\t_1+R_1\t_2+\bomega(R_1,R_2)+R_1R_2 \t_3,\bomega(R_1R_2,R_3)),
\end{multline}
\begin{multline}
B=\sigma(\t_1,R_1\t_2+R_1R_2\t_3+R_1\bomega(R_2,R_3))\\\sigma(\t_1+R_1\t_2+R_1R_2\t_3+R_1\bomega(R_2,R_3),\bomega(R_1,R_2R_3))\sigma(\t_2,R_2\t_3)\\\sigma(\t_2+R_2\t_3,\bomega(R_2,R_3))
\end{multline}
and
\begin{equation}
C=g^{-1}(R_1\bomega(R_2,R_3),R_1)\frac{\sigma(R_1\bomega(R_2,R_3),\bomega(R_1,R_2R_3))}{\sigma(\bomega(R_1,R_2),\bomega(R_1R_2,R_3))}.
\end{equation}
Note that (7) has been used to eliminate $\alpha$ from the equation.

We now consider the specification with $R_2=E$. Then, $C=1$, and
\begin{equation}
\frac{A}{B}=\frac{\sigma(\t_1,R_1\t_2)\sigma(\t_1+R_1\t_2,R_1\t_3)}{\sigma(\t_1,R_1(\t_2+\t_3))\sigma(\t_2,\t_3)}.
\end{equation}
Because $\sigma$ is a factor system for the translational subgroup, it satisfies the identity,
\begin{equation}
\sigma(\t_1,R_1\t_2)\sigma(\t_1+R_1\t_2,R_1\t_3)=\sigma(\t_1,R_1(\t_2+\t_3))\sigma(R_1\t_2,R_1\t_3).
\end{equation}
Then, 
\begin{equation}
\frac{A}{B}C=\frac{\sigma(R_1\t_2,R_1\t_3)}{\sigma(\t_2,\t_3)}.
\end{equation}
The left-hand side of Eq.~\eqref{sigma_Step1} can be easily simplified, and  Eq.~\eqref{sigma_Step1} is converted to be
\begin{equation}
\frac{g(R_1\t_2,R_1)g(R_1\t_3,R_1)}{g(R_1\t_2+R_1\t_3,R)} = \frac{\sigma(R_1\t_2,R_1\t_3)}{\sigma(\t_2,\t_3)}.
\end{equation}
Then, by redefining the arguments, we see the above equation is equivalent to (5) in the main text.

\subsubsection{Twisted covariant equation for $g$}
To prove (6) in the main text, we specialize in the case with $R_3=E$. Then, $C=1$ and equation \eqref{sigma_Step1} is reduced to be
\begin{equation}
\begin{split}
&\frac{g(R_1\t_2,R_1)g(R_1R_2\t_3,R_1R_2)}{g(R_1\t_2+R_1R_2\t_3,R_1)g(R_2\t_3,R_2)}\\
=&\frac{\sigma(\t_1,R_1\t_2)\sigma(\t_1+R_1\t_2,\bomega(R_1,R_2))\sigma(\t_1+R_1\t_2+\bomega(R_1,R_2),R_1R_2\t_3)}{\sigma(\t_1,R_1\t_2+R_1R_2\t_3)\sigma(\t_1+R_1\t_2+R_1R_2\t_3,\bomega(R_1,R_2))\sigma(\t_2,R_2\t_3)}.
\end{split}
\end{equation}
From Eq.~(5), 
\begin{equation}
\frac{g(R_1\t_2+R_1R_2\t_3,R_1)}{g(R_1\t_2,R_1)g(R_1R_2\t_3,R_1)}=\frac{\sigma(\t_2,R_2\t_3)}{\sigma(R_1\t_2,R_1R_2\t_3)}.
\end{equation}
The above two equations lead to
\begin{equation}
\begin{split}
&\frac{g(R_1R_2\t_3,R_1R_2)}{g(R_1R_2\t_3,R_1)g(R_2\t_3,R_2)}\\
=& \frac{\sigma(\t_1,R_1\t_2)\sigma(\t_1+R_1\t_2,\bomega(R_1,R_2))\sigma(\t_1+R_1\t_2+\bomega(R_1,R_2),R_1R_2\t_3)}{\sigma(\t_1,R_1\t_2+R_1R_2\t_3)\sigma(\t_1+R_1\t_2+R_1R_2\t_3,\bomega(R_1,R_2))\sigma(R_1\t_2,R_1R_2\t_3)}.
\end{split}
\end{equation}
We can use the cocycle equation for $\sigma$ to simplify the right hand side of the above equation as
\begin{equation}
\frac{g(R_1R_2\t_3,R_1R_2)}{g(R_1R_2\t_3,R_1)g(R_2\t_3,R_2)}=\frac{\sigma(\bomega(R_1,R_2),R_1R_2\t_3)}{\sigma(R_1R_2\t_3,\bomega(R_1,R_2))}.
\end{equation}
Redefining $\t=R_1R_2\t_3$ in the above equation, we arrive at (6) in the main text.

\section{(co)homology groups for Wallpaper group $pg$}
\begin{figure}
	\includegraphics[scale=0.4]{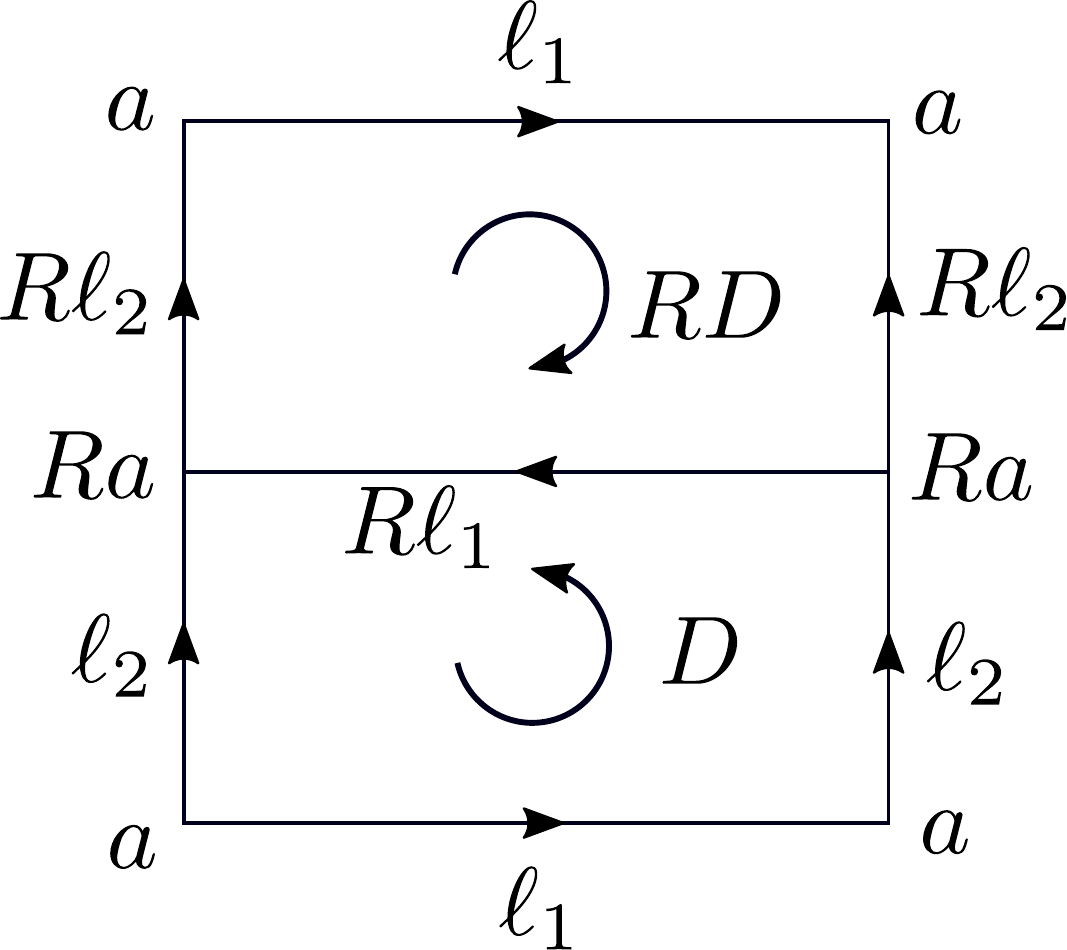}
	\caption{\label{pg-cell}The cellular structure of $T^2$ stable under the glide reflection. }
\end{figure}
The cellular structure of $T^2$ stable under the glide reflection $R$ is given in Fig.\ref{pg-cell}, leading to the chain complex $C_*$:
\begin{equation}
0\rightarrow \Z_D\oplus\Z_{RD}\xrightarrow{\partial} \Z_{\ell_1}\oplus\Z_{R\ell_1}\oplus\Z_{\ell_2}\oplus\Z_{R\ell_2}\xrightarrow{\partial}\Z_{a}\oplus\Z_{Ra}\rightarrow 0.
\end{equation}
The $D_1$-action preserving boundary operator is specified by
\begin{equation}
\partial D=\ell_1+R\ell_1,\quad \partial\ell_1=0,\quad \partial \ell_2=Ra-a.
\end{equation}
The $D_1$ resolution $F_*$ over $\Z$ can be chosen as  
\begin{equation}\label{CN-resolution}
\cdots\xrightarrow{\partial_o} \Z_E\oplus\Z_R\xrightarrow{\partial_e}\Z_E\oplus\Z_R \xrightarrow{\partial_o} \Z_E\oplus\Z_R \xrightarrow{\partial_e} \Z_E\oplus\Z_R \xrightarrow{\partial_o} \Z_E\oplus\Z_R \xrightarrow{\varepsilon}\Z\rightarrow 0 
\end{equation}
where the boundary operators are specified as
\begin{equation}
\varepsilon E=\varepsilon R=1,
\end{equation}
\begin{equation}
\partial_o E=E-R,\quad \partial_o R=R-E,
\end{equation}
and
\begin{equation}
\partial_e E=E+R,\quad \partial_e R=E+R.
\end{equation}
We then use the two chain complexes with respective $D_1$ actions to construct a total chain complex $TC_*$ with $D_1$ action being factored out.
Since $F_p\cong \Z_E\oplus\Z_R$ for all $p\ge 0$, we have
\begin{equation}
F_p\otimes_G C_q\cong C_q.
\end{equation}
Hence, the chains of $TC_*$ are derived as
\begin{equation}
\begin{split}
TC_0 &=F_0\otimes C_0\cong \Z_a\oplus \Z_{Ra}\\
TC_1 &=F_1\otimes C_0\oplus F_0\otimes C_1 \cong (\Z_a\oplus \Z_{Ra})\oplus(\Z_{\ell_1}\oplus\Z_{R\ell_1}\oplus\Z_{\ell_2}\oplus\Z_{R\ell_2})\\
TC_p &=F_p\otimes C_0\oplus F_{p-1}\otimes C_1\oplus F_{p-2}\otimes C_2\\
&\cong (\Z_a\oplus \Z_{Ra})\oplus(\Z_{\ell_1}\oplus\Z_{R\ell_1}\oplus\Z_{\ell_2}\oplus\Z_{R\ell_2})\oplus (\Z_D\oplus \Z_{RD}),~p\ge 2.
\end{split}
\end{equation}
Then, the boundary operator $\partial$ of $TC_*$ is presented as follows. First, for
\begin{equation}
\partial: F_p\otimes_{C_2} C_{0}\rightarrow F_{p-1}\otimes_{C_2} C_{0}
\end{equation}
the boundary operator is given by
\begin{equation}
\partial a=\partial_p I\otimes a =\begin{cases}
0 & p=0\\
-a+Ra & p~ odd\\
a+ Ra & p>0~\mathrm{and}~even
\end{cases}
\end{equation}
\begin{equation}
\partial Ra=\partial_p I\otimes Ra =\begin{cases}
0 & p=0\\
a-Ra & p~ odd\\
a+ Ra & p>0~\mathrm{and}~even
\end{cases}
\end{equation}
Second, for
\begin{equation}
\partial: F_p\otimes_{C_2} C_{1}\rightarrow F_{p-1}\otimes_{C_2} C_{1}\oplus F_p\otimes_{C_2} C_{0}
\end{equation}
the boundary operator is given by
\begin{equation}
\partial(\ell_1)=\partial_p I\otimes \ell_1+(-1)^p I\otimes \partial \ell_1=\begin{cases}
0 & p=0\\
-\ell_1+R\ell_1 & p~odd\\
\ell_1+R\ell_1  & p>0~\mathrm{and}~even
\end{cases}
\end{equation}
\begin{equation}
\partial(R\ell_1)=\partial_p I\otimes R\ell_1+(-1)^p I\otimes \partial R\ell_1=\begin{cases}
0 & p=0\\
\ell_1-R\ell_1 & p~odd\\
\ell_1+R\ell_1  & p>0~\mathrm{and}~even
\end{cases}
\end{equation}
\begin{equation}
\partial(\ell_2)=\partial_p I\otimes \ell_2+(-1)^p I\otimes \partial \ell_2=\begin{cases}
-a+Ra & p=0\\
a-Ra-\ell_2+R\ell_2 & p~odd\\
-a+Ra+\ell_2+R\ell_2 & p>0~\mathrm{and}~even
\end{cases}
\end{equation}
\begin{equation}
\partial(R\ell_2)=\partial_p I\otimes R\ell_2+(-1)^p I\otimes \partial R\ell_2=\begin{cases}
a-Ra & p=0\\
-a+Ra+\ell_2-R\ell_2 & p~odd\\
a-Ra+\ell_2+R\ell_2 & p>0~\mathrm{and}~even
\end{cases}
\end{equation}
Third, for
\begin{equation}
\partial: F_p\otimes C_{2}\rightarrow F_{p-1}\otimes C_{2}\oplus F_p\otimes C_1
\end{equation}
the boundary operator is given by
\begin{equation}
\partial D=\partial_p I\otimes D+(-1)^p F_p\otimes \partial D=\begin{cases}
\ell_1+R\ell_1 & p=0\\
-\ell_1-R\ell_1-D+RD & p~odd\\
\ell_1+R\ell_1+D+RD & p>0~\mathrm{and}~even
\end{cases}
\end{equation}
\begin{equation}
\partial RD=\partial_p I\otimes RD+(-1)^p F_p\otimes \partial RD=\begin{cases}
\ell_1+R\ell_1 & p=0\\
-\ell_1-R\ell_1+D-RD & p~odd\\
\ell_1+R\ell_1+D+RD & p>0~\mathrm{and}~even
\end{cases}
\end{equation}
Note that for any cell $X$, the boundary operator preserves the group action with $\partial RX=R\partial X$. 
We thus present the boundary operators at each degree in the matrix form below.
\begin{equation}
\partial_1=\begin{bmatrix}
-1 & 1   &  0 & 0 & -1 & 1\\
1  & -1  &  0 & 0 &  1 & -1
\end{bmatrix}
\end{equation}
\begin{equation}
\partial_2=\begin{bmatrix}
1 & 1 & 0  & 0  & 1  & -1 & 0 & 0 \\
1 & 1 & 0  & 0  & -1 & 1  & 0 & 0 \\
0 & 0 & -1 & 1  & 0  & 0  & 1 & 1 \\
0 & 0 & 1  & -1 & 0  & 0  & 1 & 1 \\
0 & 0 & 0  & 0  & -1 & 1  & 0 & 0 \\
0 & 0 & 0  & 0  & 1  & -1 & 0 & 0
\end{bmatrix}
\end{equation}
Boundary operators with degree greater than $2$ are given by
\begin{equation}
\partial_{2n-1}=\begin{bmatrix}
-1 & 1  & 0  & 0  & -1 & 1  & 0  & 0  \\
1  & -1 & 0  & 0  &  1 & -1 & 0  & 0  \\
0  & 0  & 1  & 1  &  0 & 0  & -1 & -1 \\
0  & 0  & 1  & 1  &  0 & 0  & -1 & -1 \\
0  & 0  & 0  & 0  &  1 & 1  & 0  & 0  \\
0  & 0  & 0  & 0  &  1 & 1  & 0  & 0  \\
0  & 0  & 0  & 0  &  0 & 0  & -1 & 1  \\
0  & 0  & 0  & 0  &  0 & 0  & 1  & -1
\end{bmatrix},
\end{equation}
\begin{equation}
\partial_{2n}=\begin{bmatrix}
1  & 1  & 0  & 0  & 1  & -1 & 0  & 0 \\
1  & 1  & 0  & 0  & -1 & 1  & 0  & 0 \\
0  & 0  & -1 & 1  & 0  & 0  & 1  & 1 \\
0  & 0  & 1  & -1 & 0  & 0  & 1  & 1 \\
0  & 0  & 0  & 0  & -1 & 1  & 0  & 0 \\
0  & 0  & 0  & 0  & 1  & -1 & 0  & 0 \\
0  & 0  & 0  & 0  & 0  & 0  & 1  & 1 \\
0  & 0  & 0  & 0  & 0  & 0  & 1  & 1 
\end{bmatrix}
\end{equation}
with $n>1$.
The Smith normal forms of the boundary matrices can be calculated by Wolfram Mathematica, which are given below.
For $\partial_{1,2}$,
\begin{equation}
\mathrm{Sm}(\partial_1)=\begin{bmatrix}
1 & 0_{1\times 5} \\
0 & 0_{1\times 5}
\end{bmatrix},
\end{equation}
\begin{equation}
\mathrm{Sm}(\partial_2)=\begin{bmatrix}
1_3 & 0_{3\times 1} & 0_{3\times 4} \\
0_{1\times 3} & 2 &  0_{1\times 4}  \\
0_{2\times 3} & 0_{2\times 1} & 0_{2\times 4}
\end{bmatrix}.
\end{equation}
For $n>2$
\begin{equation}
\mathrm{Sm}(\partial_n)=\begin{bmatrix}
1_4 & 0_{4\times 4}\\
0_{4\times 4} & 0_{4\times 4}
\end{bmatrix}.
\end{equation}
Thus, we conclude our calculation that
\begin{equation}
H_0(pg)\cong\Z,\quad H_1(pg)\cong\Z\oplus\Z_2, \quad H_n(pg)=0~ \mathrm{for}~ n>1.
\end{equation}
The systematic algebraic calculation agrees with the topological result $H_p(\Gamma_{pg})=H_p(K)$, with $K$ the Klein bottle. Since $\Gamma_{pg}$ acts freely on $\R^2$, the classifying space of $\Gamma_{pg}$ is $B_{\Gamma_{pg}}=\R^2/\Gamma_{pg}$ homeomorphic to $K$.

With the homology groups, we can derive the cohomology groups according to the universal coefficient theorem:
\begin{equation}
\mathcal{H}^n(X,A)\cong \Hom(\mathcal{H}_n(X),A)\oplus \Ext(\mathcal{H}_{n-1}(X),A).
\end{equation}
Thus,
\begin{equation}
H^0(pg,A)\cong A,\quad H^1(pg,A)\cong A\oplus \Z_2.
\end{equation}
and 
\begin{equation}
H^2(pg,U(1))\cong 0,\quad H^2(pg,\Z_2)=0.
\end{equation}

\section{Factor systems for wallpaper group $pg$}

The wallpaper group $pg$ is the simplest nonsymmorphic group with only one generator, a glide reflection. The wallpaper group operators on a rectangular lattice. We choose the unit translational vectors as $\bm{e}_a=(a,0)$ and $\bm{e}_b=(0,b)$. The glide reflection is chosen as $\{0|M_1\}$, where $M_1$ is the reflection with respect to the $x$-axis, and $\btau(M)=\e_a/2$. Therefore, the action of $\{0|M\}$ on the 2D plane $\R^2$ is given by
\begin{equation}
\{0|M\} \begin{bmatrix}
x \\ y
\end{bmatrix}=\begin{bmatrix}
x+{a}/{2} \\
-y
\end{bmatrix}.
\end{equation}
Accordingly,
\begin{equation}
\omega(M,M)=\bm{e}_a.
\end{equation}
It is straightforward to check that $\sigma^\mu(M\t_1,M\t_2)=\sigma^\mu(\t_1,\t_2)$ for both $\mu=0,1$. Hence, $g(\t_1+\t_2,R)=g(\t_1,R)g(\t_2,R)$, and the solution for $g$ takes the general form,
\begin{equation}
g(\t,R)=(-1)^{\k(R)\cdot\t},
\end{equation}
where $\k$ maps $D_1=\{E,M\}$ into inversion-invariant points in the Brillouin zone. For Eq.~\eqref{Covariance_g}, the left-hand side is constantly equal to $1$ with the solution for $g$ above. This requires that the factor system $\sigma$ for the translational subgroup be trivial,
\begin{equation}
\sigma=\sigma^0.
\end{equation}
For nontrivial $\sigma^1$, the right-hand side of Eq.~6 in the main text with $R_1=R_2=M$ is equal to 
\begin{equation}
\frac{\sigma(\omega(M,M),\t)}{\sigma(\t,\omega(M,M))}=(-1)^{t^b},
\end{equation}
which contradicts with the left-hand side.

According to Eq.~(6) in the main text with right-hand side being 1, the only nontrivial equation for $\k$ is
\begin{equation}
\k(E)=\k(M)+M\k(M) \mod \G,
\end{equation}
which leads to two independent solutions $\k^{(1,0)}(M)=(1/a,0)^T$ and $\k^{(0,1)}(M)=(0,1/b)^T$.
We further consider the case of $R_1=R_2=R_3=M$ for Eq.~(7) in the main text, which leads to
\begin{equation}
1=(-1)^{k^a(M)}.
\end{equation}
Hence, $\k^{(1,0)}$ should be dropped out, and only $\k^{(0,1)}$ is consistent with Eqs.~(6) and (7). Then, we have two solutions for $\k$ given by
\begin{equation}
\k^\nu(M)=(0,\nu)^T
\end{equation}
with $\nu=0,1$, forming an abelian group $\Z_2$.  Moreover, for both $\k^\nu$, $\alpha$ satisfies the homogeneous 2-cocyle equation with solutions forming $\Z_2$.

From above derivations, we obtain four factor systems as an abelian group $\Z_2\times\Z_2$. But the second cohomology is $H^{2}(pg,\Z_2)\cong \Z_2$. Therefore, only two of the four factor systems are not equivalent. We now consider the function,
\begin{equation}
\chi(\{\t|R\})=(-1)^{t^a},
\end{equation}
which leads to the coboundary
\begin{equation}
\frac{\chi(\{\t_1|R_1\}\{\t_2|R_2\})}{\chi(\{\t_1|R_1\})\chi(\{\t_2|R_2\})}=(-1)^{\omega^{a}(R_1,R_2)}.
\end{equation}
Then, it is immediately realized that the $\Z_2$ component from $\alpha$ can be divided out. Hence, the two non-equivalent factor systems are only specified by $g$ as 
\begin{equation}
\nu(\{\t_1|R_1\},\{\t_2|R_2\})=g_\nu(R_1\t_2,R_1)=(-1)^{\k^\nu(R_1)\cdot R_1\t_2}.
\end{equation}

Concluding, there is only one nontrivial factor system class for $pg$, which is represented by
\begin{equation}
\nu(\{\t_1|R_1\},\{\t_2|R_2\})=\begin{cases}
1 & R_1=E\\
(-1)^{t^b_2} & R_1=M
\end{cases}.
\end{equation}

\section{Nonsymmorphic wallpaper groups}
\begin{table}
	\begin{tabular}{c|c|c|c|c}
		\hline
		Wallpaper group &~~~$pg$~~~&~~~$pmg$~~~&~~~$pgg$~~~&~~~$p4g$~~~ \\
		\hline
		Point Group	    & $D_1$  & $D_2$    & $D_2$    & $D_4$ \\
		$H^2(P,\Z_2)$ & $\Z_2$ & $\Z_2^3$ & $\Z_2^3$ & $\Z_2^3$ \\
		\hline
		$g$           & $\Z_2$ & $\Z_2^2$ & $\Z_2$ & $\Z$ \\
		\hline
		$H^2(G,\Z_2)$ & $\Z_2$ & $\Z_2^4$ & $\Z_2^2$ & $\Z_2^3$ \\
		\hline
	\end{tabular}
	\caption{The cohomology groups for nonsymmorphic wallpaper groups. Here, $\Z_2^n$ denotes the direct product of $n$ copies of $\Z_2$.\label{tab:Nonsym-Wallpaper}}
\end{table}
With $\sigma=1$, we then solve the homogeneous Eq.~(6) in the main text, and the solutions for $g$ form abelian groups, which are listed in Tab.\ref{tab:Nonsym-Wallpaper}. As aforementioned, possibilities of $\alpha$ is also characterized by $H^{2}(P,\Z_2)$. We observe from Tab.\ref{tab:Nonsym-Wallpaper} that the direct sum of the abelian group of $g$'s and $H^2(P,\Z_2)$ is greater than $H^2(G,\Z_2)$. This is because some pairs $(g,\alpha)$ become equivalent by considering the equivalence relations (9) and (10) in the main text imposed by $\psi(\k)=(-1)^{\K_a\cdot\t}$.

\end{document}